\documentclass[aps,prb,twocolumn,amsmath,amssymb,showpacs,superscriptaddress]{revtex4-1}

\usepackage{graphicx}
\usepackage{amsmath,amssymb,amsfonts}
\usepackage{textcomp}
\usepackage{hyperref}
\usepackage{gensymb}
\usepackage{verbatim}
\usepackage{amsmath}
\hypersetup{breaklinks=true,colorlinks=true,urlcolor=black}
\usepackage{color}
\usepackage{soul}
\DeclareGraphicsExtensions{.jpg,.pdf,.png,.eps}

\begin{document}
	\title{Pressure-induced multiple phase transformations of the BaBi$_3$ superconductor}
	\author{Li Xiang}
	\affiliation{Ames Laboratory, Iowa State University, Ames, Iowa 50011, USA}
	\affiliation{Department of Physics and Astronomy, Iowa State University, Ames, Iowa 50011, USA}
	\email[]{ives@iastate.edu}
	\author{Raquel A. Ribeiro}
	\affiliation{Department of Physics and Astronomy, Iowa State University, Ames, Iowa 50011, USA}
	\affiliation{CCNH, Universidade Federal do ABC (UFABC), Santo Andr\'e, SP, Brazil }
	\author{Udhara S. Kaluarachchi}
	\affiliation{Ames Laboratory, Iowa State University, Ames, Iowa 50011, USA}
	\affiliation{Department of Physics and Astronomy, Iowa State University, Ames, Iowa 50011, USA}
	\author{Elena Gati}
	\affiliation{Ames Laboratory, Iowa State University, Ames, Iowa 50011, USA}
	\author{Manh Cuong Nguyen}
	\affiliation{Ames Laboratory, Iowa State University, Ames, Iowa 50011, USA}
	\author{Cai-Zhuang Wang}
	\affiliation{Ames Laboratory, Iowa State University, Ames, Iowa 50011, USA}
	\author{Kai-Ming Ho}
	\affiliation{Ames Laboratory, Iowa State University, Ames, Iowa 50011, USA}
	\author{Sergey L. Bud'ko}
	\affiliation{Ames Laboratory, Iowa State University, Ames, Iowa 50011, USA}
	\affiliation{Department of Physics and Astronomy, Iowa State University, Ames, Iowa 50011, USA}
	\author{Paul C. Canfield}
	\affiliation{Ames Laboratory, Iowa State University, Ames, Iowa 50011, USA}
	\affiliation{Department of Physics and Astronomy, Iowa State University, Ames, Iowa 50011, USA}
	\email[]{canfield@ameslab.gov}
	
\date{\today}

\begin{abstract}
Measurements of temperature-dependent resistance and magnetization under hydrostatic pressures up to 2.13 GPa are reported for single-crystalline, superconducting BaBi$_3$. A temperature - pressure phase diagram is determined and the results suggest three different superconducting phases ${\alpha}$, $\beta$, and $\gamma$ in the studied pressure range. We further show that occurrence of the three superconducting phases is intuitively linked to phase transitions at higher temperature which are likely first order in nature. $T_p$, which separates phase ${\alpha}$ from $\beta$ and $\gamma$, is associated with an abrupt resistance change as pressure is increased from 0.27 GPa to 0.33 GPa. Above 0.33 GPa, an "S-shape" anomaly in the temperature-dependent resistance curve, $T_\text S$, is observed and associated with the transition between the $\beta$ and $\gamma$ phases. Further increasing of pressure above 1.05 GPa suppresses this transition and BaBi$_3$ stays in $\gamma$ phase over the whole investigated temperature range. These high-temperature anomalies are likely related to structural degrees of freedom. With the ${\alpha}$ phase being the ambient-pressure tetragonal structure ($P4/mmm$), our first-principle calculations suggest the $\beta$ phase to be cubic structure ($Pm-3m$) and the $\gamma$ phase to be a distorted tetragonal structure where the Bi atoms are moved out of the face-centered position. Finally, an analysis of the evolution of the superconducting upper critical field with pressure further confirms these transitions in the superconducting state and suggests a possible change of band structure or a Lifshitz transition near 1.54 GPa in $\gamma$ phase. Given the large atomic numbers of both Ba and Bi, our results establish BaBi$_3$ as a good candidate for the study of the interplay of structure with superconductivity in the presence of strong spin-orbit coupling.
\end{abstract}
	
\maketitle 
	
\section{Introduction}
Materials with strong spin-orbit coupling have recently received a lot of attention as they are argued to be hosts for novel topological phases, such as topological insulators or topological superconductors\cite{Hasan2010,Qi2011,Hor2010}. Among them, Bi-based compounds are among the most investigated for their strong spin-orbit coupling due to Bi-$6p$ electrons\cite{Anna2013}. For example, the compounds Bi$_2$$X_3$ ($X$ = Se, Te) are suggested to be topological insulators\cite{Xia2009,Chen2009}.

Another Bi-rich family of compounds $A$Bi$_3$ ($A$ = Sr, Ba and La) has attracted attention lately as these materials are superconductors. Polycrystalline $A$Bi$_3$ compounds with $A$ = Sr and Ba were first reported to be superconductors by Matthias and Hulm in 1952\cite{Matthias1952PR}. Later on, single crystals of SrBi$_3$ and BaBi$_3$ were synthesized using the Bi self-flux method\cite{Canfield1992} by various research groups and were reported to have superconducting transition temperatures $T_\text c$ of 5.75 K and 5.9 K, respectively\cite{Kakihana2015,Haldolaarachchige2014,Jha2017}. Furthermore, Na substitution for Sr in SrBi$_3$ increases $T_\text c$ to 9.0 K\cite{Iyo2015}. Polycrystalline LaBi$_3$ was synthesized more recently by utilizing a high-pressure technique\cite{Kinjo2016} and reported to have a $T_\text c$ of 7.3 K. Among the three $A$Bi$_3$ compounds, SrBi$_3$ and LaBi$_3$ crystallize in the AuCu$_3$-type cubic structure ($Pm-3m$), whereas BaBi$_3$ crystallizes in tetragonal structure ($P4/mmm$) with only a small difference in $a$ and $c$ lattice parameters ($a$ = 5.06(1) {{\AA}} and $c$ = 5.13(2) {{\AA}})\cite{Haldolaarachchige2014,Shao2016,Jha2017}. Importantly, for all three $A$Bi$_3$ compounds, spin-orbit coupling (SOC) is suggested to play an significant role in the superconductivity\cite{Shao2016}, making the $A$Bi$_3$ compounds potential platforms for the realization of topological superconductivity.

Further insight into the nature of the superconductivity can be obtained by studying the system's response to hydrostatic pressure. As a tuning parameter, pressure is considered clean compared to substitution since it does not induce extra chemical disorder into the systems. It has been proven to be very useful in terms of tuning the ground state in many systems\cite{Schilling1979,S.Schilling2001,Lorenz2005,Jackson2005}, such as Fe-based superconductors\cite{Colombier2009,Chu2009,Taufour2014,Xiang2017} and quantum-critical materials\cite{Saiga2008JPSJ,Matsubayashi2010,Kim2013,Kaluarachchi2017natcomms}. Earlier studies of the effect of hydrostatic pressure on $A$Bi$_3$ revealed that, for LaBi$_3$ and SrBi$_3$, pressure linearly suppresses $T_\text c$ up to 1.55 GPa and 0.81 GPa, respectively\cite{Kinjo2016,Jha2017}. Interestingly, BaBi$_3$ was shown to exhibit a double-transition feature in the temperature-dependent magnetization curves for pressures above 0.3 GPa\cite{Jha2017}. However, the origin and nature of the feature has not been studied in greater detail up to now.

In this work, we present a detailed pressure study on BaBi$_3$ by utilizing both resistance and magnetization measurements. Our data reproduce the multiple superconducting transitions in an intermediate pressure region for  0.33 GPa $\leqslant p \leqslant$ 1.05 GPa, whereas only a single sharp transition is revealed for $p \leqslant$ 0.27 GPa and $p \geqslant$ 1.27 GPa. The magnetization measurements confirm that superconductivity is not filamentary, but pressure stabilized phases. In addition, our data sets reveal a series of so far undetected high-temperature phase transitions.

From these data sets, we determine a temperature-pressure ($T-p$) phase diagram which highlights the existence of three phases (each superconducting at low temperatures) in BaBi$_3$. We argue that the high-temperature anomalies, which are likely first order in nature, are related to structural degrees of freedom. Our first-principle calculations support that several structures are close in energy for BaBi$_3$ and allow us to infer the possible pressure stabilized structures. Our results establish BaBi$_3$ as an interesting system to study the interplay of superconductivity and structural degrees of freedom in the presence of strong SOC.

\section{Experimental details}
\subsection{Experimental details}
Single crystals of mm sizes BaBi$_3$ (inset of Fig. \ref{fig1_RT} (a)) were grown through a Bi self-flux technique\cite{Canfield1992,Jha2017} with the help of a frit-disc crucible set\cite{Canfield2016a}. The ac resistance measurement under pressure was performed in a Quantum Design Physical Property Measurement System (PPMS) using a 1 mA excitation with frequency of 17 Hz, on cooling and warming at a rate of $\pm$0.25 K/min. A standard four-contact configuration was used. Contacts were made by DuPont 4929N silver paint inside a N$_2$ glove box due to the air sensitivity of the compound. The magnetic field was applied perpendicular to the current direction. A Be-Cu/Ni-Cr-Al hybrid piston-cylinder cell, similar to the one described in Ref. \onlinecite{Budko1984}, was used to apply pressure. Pressure values at low temperature were inferred from the $T_{c}(p)$ of lead\cite{Bireckoven1988}. Pressure values at higher temperature were estimated by linear interpolation between the room-temperature pressure $p_{300\textrm{K}}$ and low-temperature pressure $p_{T\leq90\textrm{K}}$ values\cite{Thompson1984,Torikachvili2015,Lamichhane2018}. $p_{300\textrm{K}}$ values were inferred from the $300$ K resistivity ratio $\rho(p)/\rho(0\,\textrm{GPa})$ of lead\cite{Eiling1981} and $p_{T\leq90\textrm{K}}$ values were inferred from the $T_{c}(p)$ of lead\cite{Bireckoven1988}. Good hydrostatic conditions were achieved by using a 4:6 mixture of light mineral oil:n-pentane as pressure medium, which solidifies, at room temperature, in the range $3-4$ GPa, i.e., well above our maximum pressure\cite{Budko1984,Kim2011,Torikachvili2015}.

Low-field (20 mT) dc magnetization measurements under pressure were performed in a Quantum Design Magnetic Property Measurement System (MPMS-3) SQUID magnetometer. A commercially-available HDM Be-Cu piston-cylinder pressure cell\cite{HDM} was used to apply pressures up to 1.2 GPa. Daphne oil 7373 was used as a pressure medium, which solidifies at ∼2.2 GPa at room temperature\cite{Yokogawa2007}, ensuring hydrostatic conditions. Slight errors in the centering of the composite Pb/BaBi$_3$ sample and pressure cell happen during the magnetization measurements, which cause the upturn features as shown in Fig. \ref{fig5_MT}. Superconducting Pb was used as a low-temperature pressure gauge\cite{Eiling1981}. Note that for both pressure cells, load was always applied at room temperature.

\subsection{Computational Methods}
To further investigate possible low energy structures of BaBi$3$, we performed a random structure search by making several hundreds of structures with different symmetries and unit cell sizes, i.e., 2, 3, 4 and 6 formula units in the unit cell. All structures were then fully relaxed by density functional theory (DFT) with criterial 0.01 eV/{{\AA}} for force components and 1 kbar (0.1 GPa) for stress tensor elements. The DFT\cite{Kohn1965} calculations were performed by Vienna $Ab-initio$ Simulation Package (VASP)\cite{Kresse1996} with projector-augmented wave (PAW) pseudo-potential method\cite{Blochl1994,Kresse1999} within generalized-gradient approximation (GGA)\cite{Perdew1996}. The Monkhost-Pack scheme\cite{Monkhorst1976} was used for Brillouin zone sampling with a high quality k-point grid of 2$\pi \times$ 0.025 {{\AA}}$^{-1}$. The energy cutoff was 320 eV and spin-orbit coupling (SOC) was included in calculations.

\section{Results and discussions}
	
\begin{figure}
	\includegraphics[width=8.6cm]{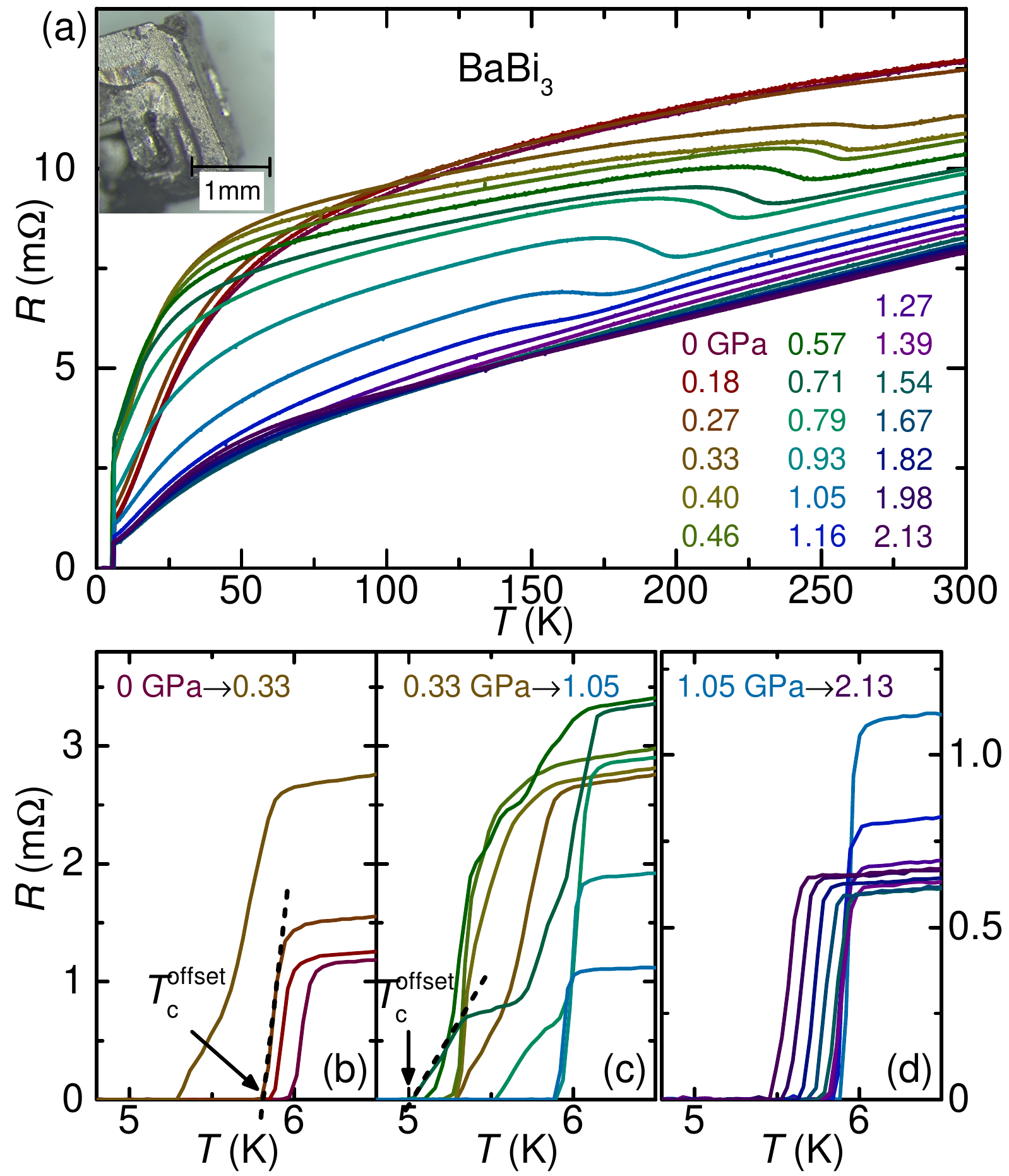}%
	\caption{(a) Evolution of the resistance with hydrostatic pressures up to 2.13 GPa. Data has been taken upon cooling; all data were taken upon increasing $p$. Pressure values in the figure legends are low-temperaure pressure values $p_{T\leq90\textrm{K}}$. (Inset) Picture a of BaBi$_3$ single crystal; (b), (c), (d) Blow ups of the low-temperature superconducting transition for three different pressure regions. The pressure regions have been chosen to represent the characteristic change of the the superconducting transition. Note that for 0.33 GPa $\leqslant p \leqslant$ 1.05 GPa in panel (c), the superconducting transition occurs in multiple steps in the $R(T)$ data. Criterion for superconducting transition temperature $T^\textrm{offset}_\textrm{c}$ is indicated by arrows in panels (b) and (c).
		\label{fig1_RT}}
\end{figure}
	
\begin{figure}
	\includegraphics[width=8.6cm]{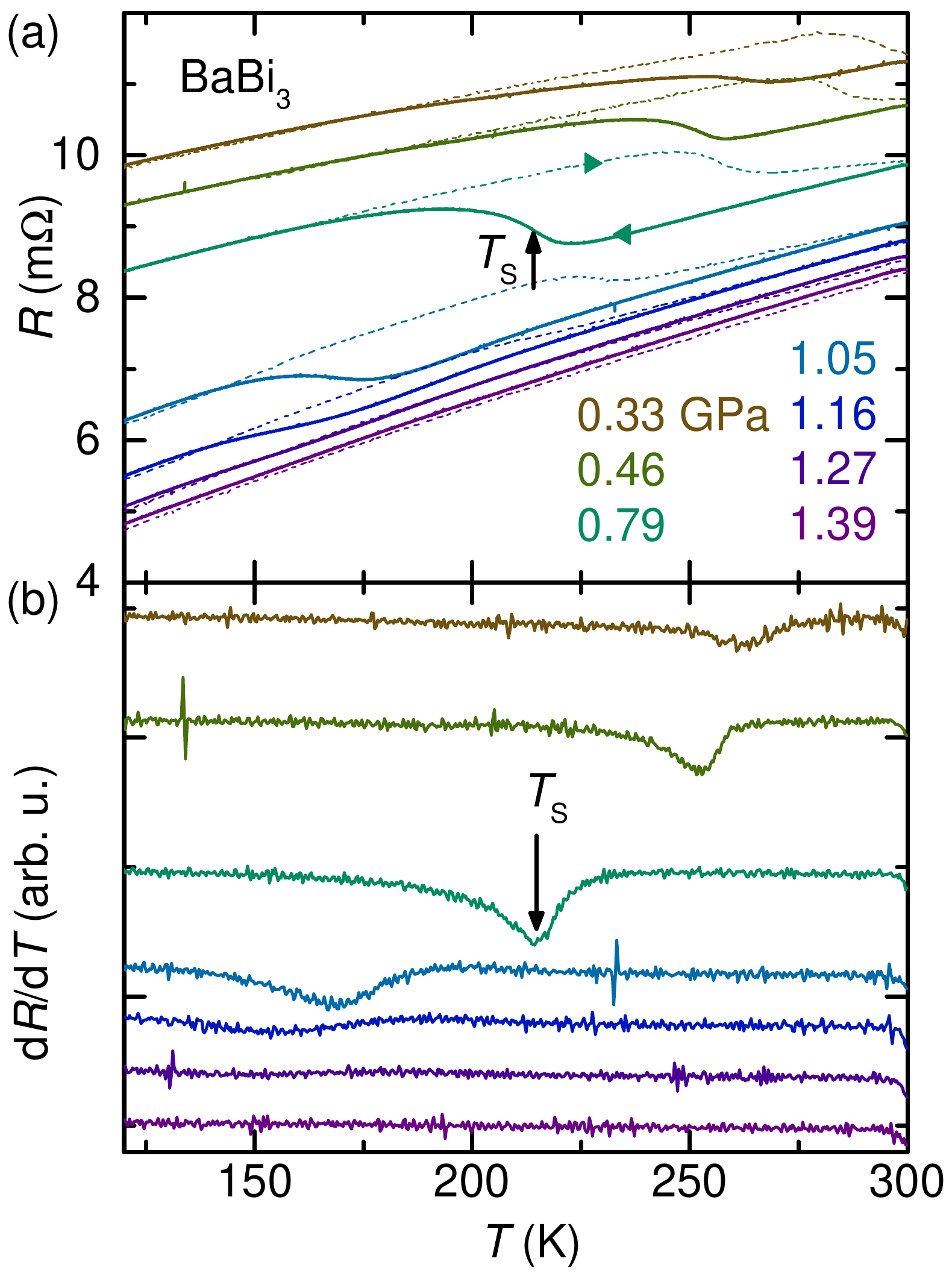}%
	\caption{(a) Temperature-dependent resistance $R$ taken on cooling (solid lines) and warming (dashed lines) for selected pressures; (b) Temperature derivative $dR/dT$ taken on cooling showing the evolution of the transition temperature $T_\textrm S$. The criterion for the determination of $T_\textrm S$ is indicated by arrow.
		\label{fig2_hysteresis}}
\end{figure}
	
\begin{figure}
	\includegraphics[width=8.6cm]{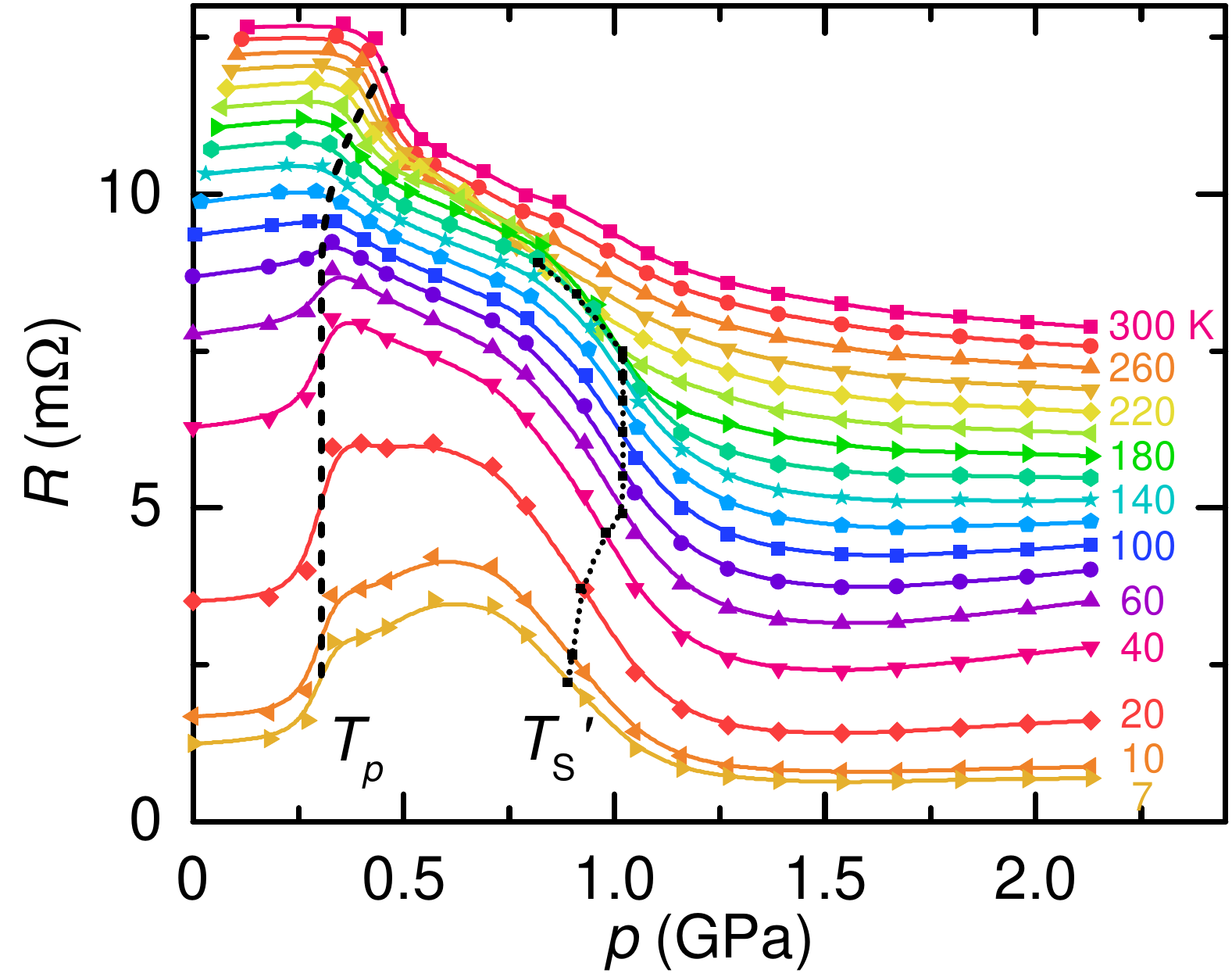}%
	\caption{The pressure dependence of resistance $R(p)$ at fixed temperatures. Pressure values are corrected for temperature-induced changes (see main text for details). The dashed line labeled $T_p$ indicates a kink-like anomaly at $p\sim$ 0.3 GPa for low temperatures and $p\sim$ 0.43 GPa for 300 K. Dotted line labeled $T_\text S'$ indicates another broad feature at $p\sim$ 1 GPa, which is discernible up to $\sim$ 220 K. Determination of $T_\text S'$, and its relation to $T_\text S$, is discussed in detail in the main text.
		\label{fig3_RP}}
\end{figure}

\begin{figure}
	\includegraphics[width=8.6cm]{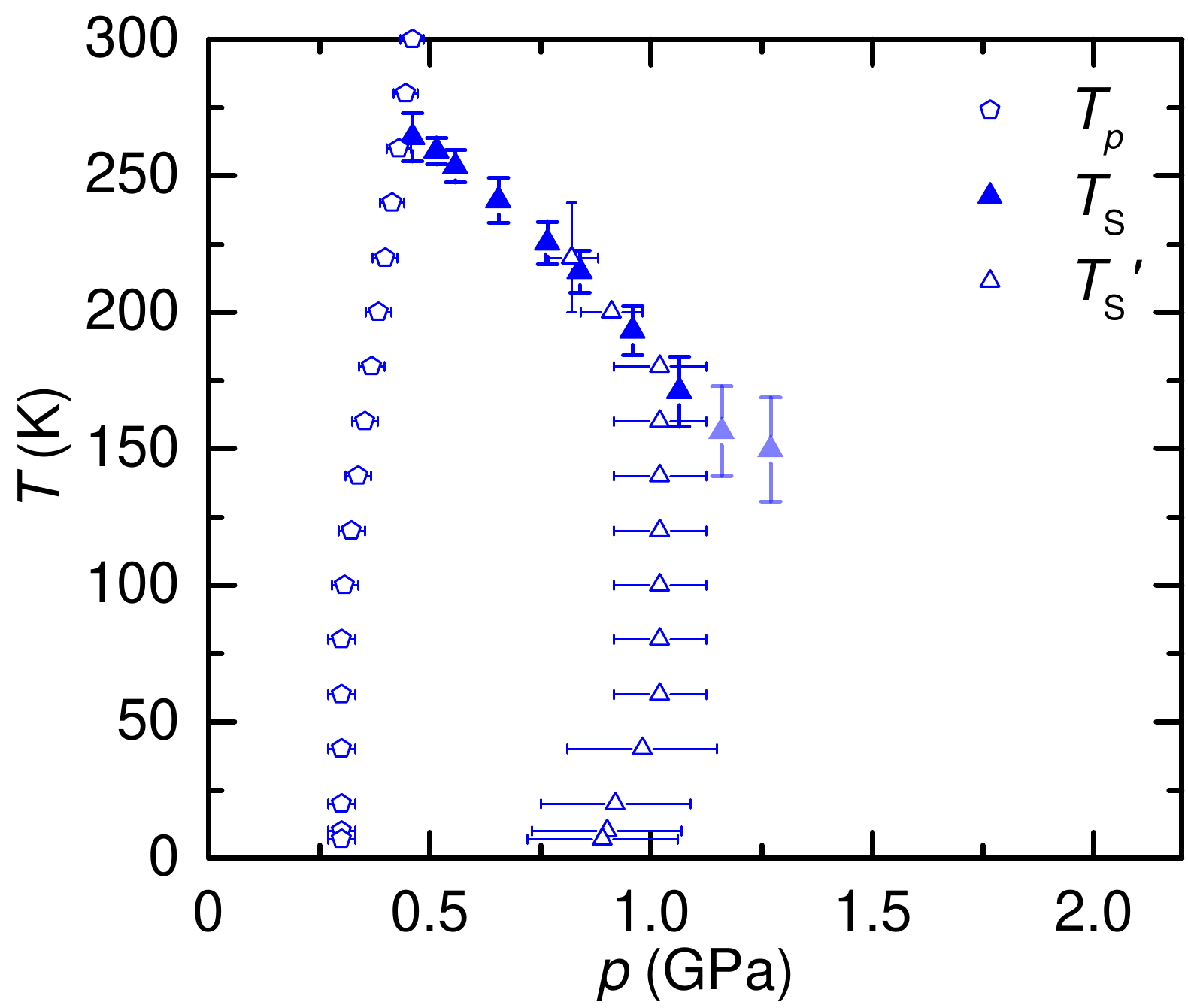}%
	\caption{Temperature-pressure phase diagram inferred from resistance measurements. The blue solid triangles represent the phase transition $T_\textrm S$ inferred from the data shown in Fig. \ref{fig2_hysteresis}. The two light-blue data points for $T_\textrm S$ are the last two, broad, barely observable features in $dR/dT$ and may not correspond to an actual transition (see main text for details). The open blue triangles and pentagons represent the pressure-induced transition $T_\text S'$ and $T_p$ inferred from the dashed and dotted lines in Fig. \ref{fig3_RP}, respectively. Superconducting transition temperature $T_\text c$ is not plotted in this phase diagram and will be discussed later.
		\label{fig4_phasediagram_transport}}
\end{figure}

Figure \ref{fig1_RT} shows the pressure  dependence of the temperature-dependent resistance $R(T)$ of BaBi$_3$. All data were taken upon increasing pressure up to 2.13 GPa. As shown in Fig. \ref{fig1_RT} (a), at 0 GPa, the resistance decreases as temperature is lowered, showing metallic behavior. Below $\sim$ 6 K, BaBi$_3$ becomes superconducting. Initially, increasing pressure suppresses the resistance value at 300 K, $R(300 \text K)$, slightly. However, when pressure is increased from 0.27 GPa to 0.33 GPa, a sudden drop in $R(300 \text K)$ is observed, and the overall behavior of temperature-dependent resistance changes as well; starting from $p=$ 0.33 GPa, a "S-shape" anomaly at $T\sim$ 250 K in the $R(T)$ curve emerges. The feature is clearly observed up to 1.05 GPa, it becomes much weaker for 1.16 GPa and 1.27 GPa and disappears for higher pressures. The transition temperature, $T_\text S$, for this anomaly is suppressed upon increasing pressure (Fig. \ref{fig2_hysteresis}). Figures \ref{fig1_RT} (b)-(d) present blow-ups of the low-temperature superconducting transition for three different pressure regions. For the low-pressure region (0 GPa $\leqslant p \leqslant$ 0.27 GPa), the superconducting transition in resistance remains sharp and single, and $T_\text c$ is suppressed by increasing pressure. For the intermediate pressure region (0.33 GPa $\leqslant p \leqslant$ 1.05 GPa), multiple steps in the superconducting transition are observed. For the high-pressure region (1.16 GPa $\leqslant p \leqslant$ 2.13 GPa), the superconducting transition becomes sharp and single again and $T_\text c$ decreases with increasing pressure as well.

In order to create a $T-p$ phase diagram, first, we focus on a more detailed analysis of the "S-shape", high-temperature feature in the intermediate pressure region. Figure \ref{fig2_hysteresis} presents the analysis of the "S-shape" anomaly in the temperature-dependent resistance curves. Figure \ref{fig2_hysteresis} (a) shows the $R(T)$ curve for the "S-shape" anomaly on both cooling (solid lines) and warming (dashed lines) for selected pressures. Clear, 10 - 25 K wide, hysteresis is observed, indicating the transition's first-order nature. The temperature derivative of the resistance, $dR/dT$, taken upon cooling, is shown in Fig. \ref{fig2_hysteresis} (b). It is clearly seen that $T_\textrm S$ is suppressed with increasing pressure. This feature is well pronounced up to 1.05 GPa, it becomes distinctly weaker for 1.16 GPa and 1.27 GPa and is not detectable anymore for higher pressures.

To follow the feature associated with the sudden change in $R(p,T=300 K)$ at $p\sim$ 0.3 GPa to lower temperatures, the pressure dependence of the resistance $R(p)$ at fixed temperatures is determined from the data in Fig. \ref{fig1_RT} (a) and presented in Fig. \ref{fig3_RP}. At low temperature, a kink-like anomaly is observed at $p\sim$ 0.3 GPa. The anomaly manifests as an increase of resistance with increasing pressure and at high temperature it manifests as a drop. This behavior reflects the crossing point of the $R(T)$ curves at $T\sim$ 110 K for $p<$0.33 GPa and $p\geqslant$ 0.33 GPa, as shown in Fig. \ref{fig1_RT}. Similar behavior has also been observed in PbTaSe$_2$ where the sudden changes in $R(p)$ is also associated with a first-order structural phase transition\cite{Kaluarachchi2017PRBa}. The anomaly moves slightly to higher pressure with increasing $T$. We stress that the pressure values given here were corrected for temperature-induced changes of the pressure (see Experimental Details). This kink-like anomaly is denoted as $T_p$ and the corresponding transition pressures have been determined from the midpoint of the jump-like change in $R(p)$. At higher pressure, another much broader transition, $T_\text S'$, is observed in Fig. \ref{fig3_RP} for $p\sim$0.8 $-$ 0.9 GPa, which exists up to $T\sim$ 220 K. To determine $T_\text S'$, for each temperature shown in the figure, the $R(p)$ data for 0.75 GPa $<p<$ 1.25 GPa was fitted using polynomial function up to the third order and the inflection point was taken as $T_\text S'$.

We summarize the position of the high-temperature anomalies observed in $R(p,T)$ in the temperature-pressure ($T-p$) phase diagram shown in Fig. \ref{fig4_phasediagram_transport}. As shown in the figure, $T_p$ (blue pentagon) is located around 0.3 GPa at low temperatures and represents the sudden change of the $R(p)$ behavior from 0.27 GPa to 0.33 GPa. The temperature of the "S-shape" anomaly, $T_\textrm S$ (blue triangle), is continuously suppressed from 264 K to 150 K by pressure. $T_\text S'$ represents the broad transition at 0.8 $\sim$ 0.9 GPa and persists up to $T\sim$ 220 K as indicated in Fig. \ref{fig3_RP}. The behavior of $T_\text c$ with pressure and its relationship with the high-temperature anomalies will be discussed later (see Fig. \ref{fig7_phasediagram} below).

\begin{figure}
	\includegraphics[width=8.6cm]{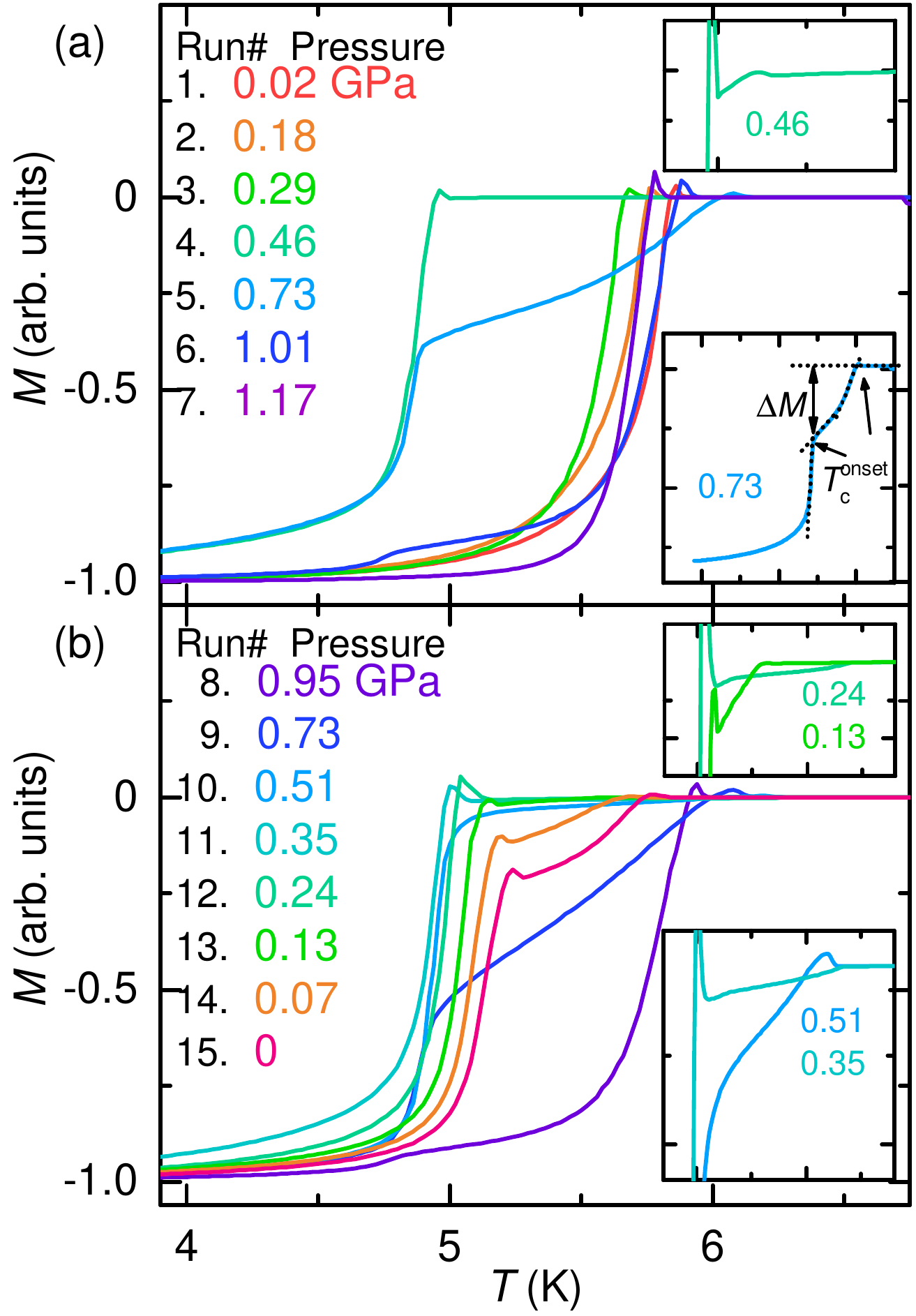}%
	\caption{Evolution of the zero-field-cooled (ZFC) magnetization $M(T)$ with (a) increasing pressure and (b) decreasing pressure in an applied field of 20 mT. The low-temperature ($T\ll T_\text c$) and higher-temperature ($T_\text c \ll T \ll T_\text {c,Pb}$) $M$ values have been set to -1 and 0, respectively, due to uncertainties involved in the determination of absolute values (see main text). The low-temperature pressure is inferred from the pressure dependence of superconducting transition of Pb (not shown). Black numbers before pressure values (Run\#) indicate the sequence of the applied pressure. Criterion for the determination of superconducting transition temperature $T^\text{onset}_\text c$ is indicated by arrows in the lower inset of panel (a). Blowups of $M(T)$ curves for several pressures in the upper inset in panel (a) and insets in panel (b) better show the double-transition feature for selected pressures. Small upturns at the onset of diamagnetism are due to slight error in the centering of the sample during measurements.
		\label{fig5_MT}}
\end{figure}
	
\begin{figure}
	\includegraphics[width=8.6cm]{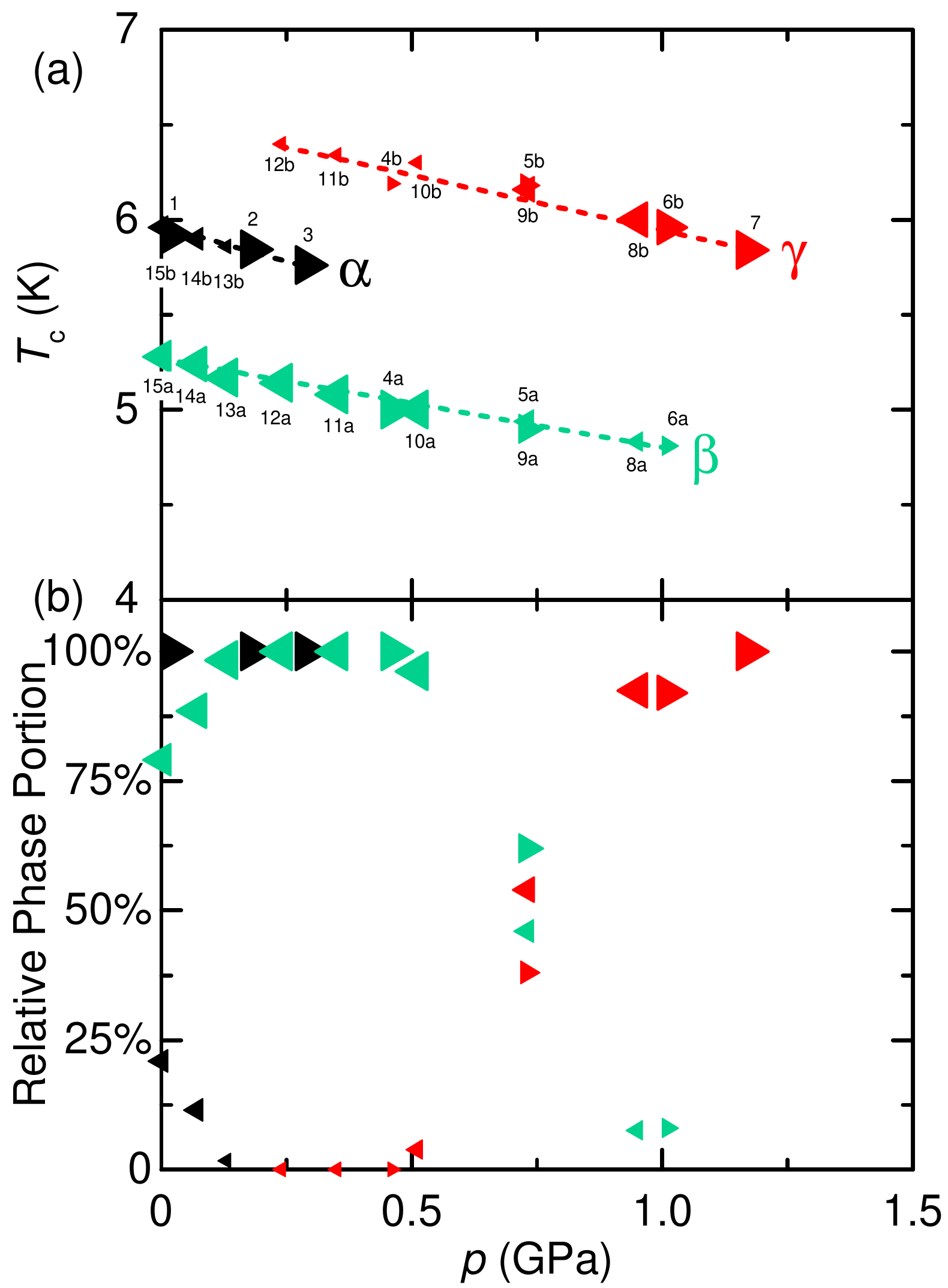}%
	\caption{(a) Temperature-pressure ($T-p$) phase diagram for superconducting $T_\text c$ inferred from magnetization measurements. The run-number associated with each data point is the same as used in Fig. \ref{fig5_MT}. The directions of the triangle data point indicate increasing ($\blacktriangleright$) and decreasing ($\blacktriangleleft$) pressure. For pressures with multiple transitions letters $a$ and $b$ are used. Three different phases ${\alpha}$ (Black), $\beta$ (Green), and $\gamma$ (Red) are suggested. Dashed lines are guides to the eye; (b) Relative phase portion as a function of pressure, as determined from magnetization measurements. Sizes of symbols in (a) and (b) are roughly proportional to the phase portion values as indicated in (b).
		\label{fig6_phasediagram_MT}}
\end{figure}

\begin{figure}
	\includegraphics[width=8.6cm]{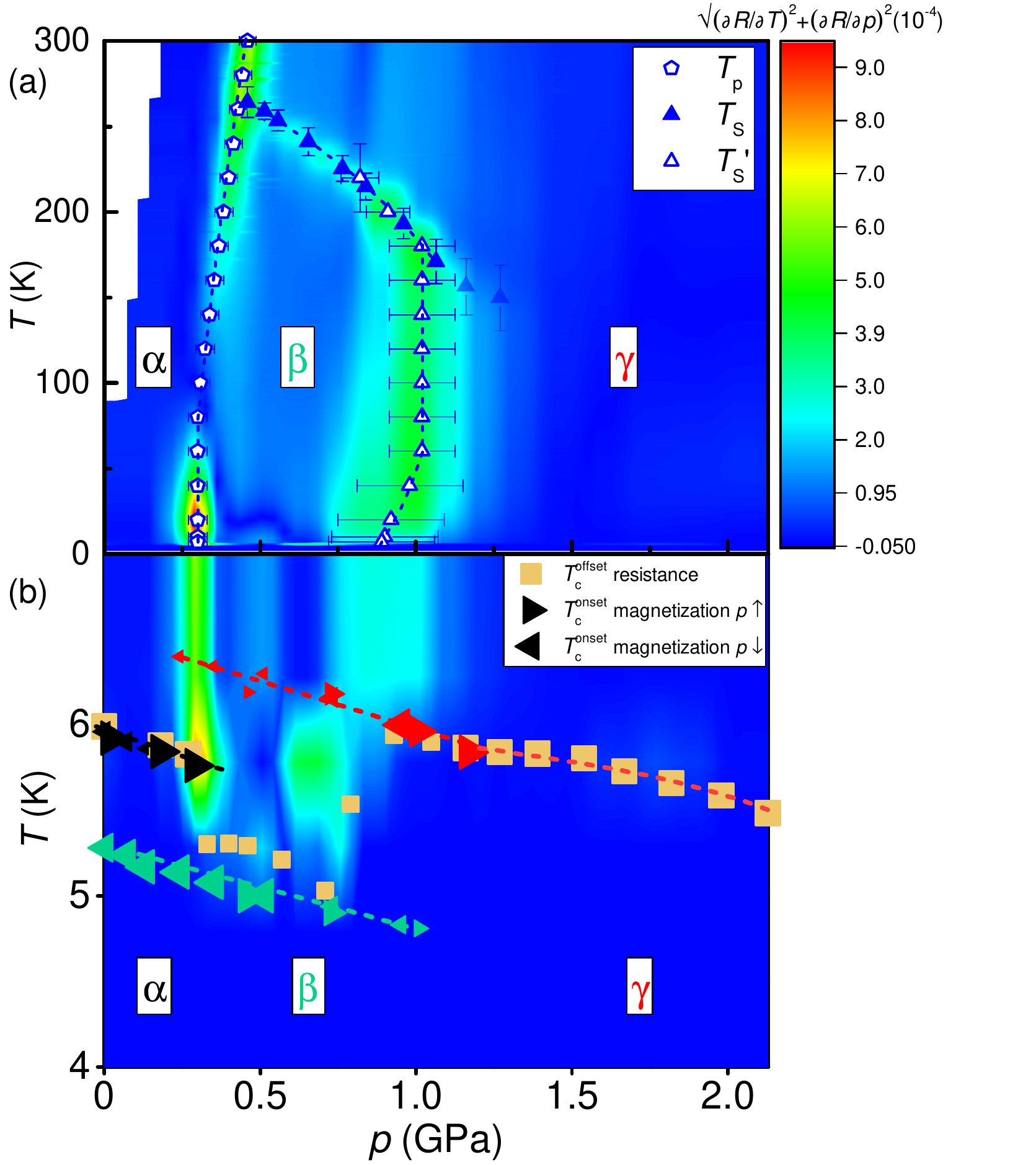}%
	\caption{(a) Color plot of surface-gradient magnitude $\sqrt{(\partial R/\partial T)^2+(\partial R/\partial p)^2}$ calculated from $R(T,p)$ data. Anomalies in the color plot coincide with the phase transitions $T_S$, $T_p$ in Fig. \ref{fig4_phasediagram_transport}. The broad transition around 1 GPa labeled as $T_\text S'$ in Fig. \ref{fig3_RP} is also revealed in this color plot. Different phases ${\alpha}$, $\beta$
	and $\gamma$ are proposed for BaBi$_3$ at different positions in the $T-p$ phase diagram; (b) Blow up of the color plot low-temperature region together with superconducting transition data from resistivity (Fig. \ref{fig4_phasediagram_transport}) and magnetization (Fig. \ref{fig6_phasediagram_MT}). Dashed lines are guides for the eye.
		\label{fig7_phasediagram}}
\end{figure}

In order to provide thermodynamic data on the superconductivity under pressure, we present, in Fig. \ref{fig5_MT}, the dependence of the zero-field-cooled (ZFC) magnetization $M(T)$ data. During the measurements, pressure was first monotonically increased from ambient pressure to 1.17 GPa, then it was decreased back to ambient pressure. The superconducting transition temperature of BaBi$_3$ is inferred from the onset of diamagnetism which is visible in all data sets under $p$, demonstrating the bulk nature of superconductivity in the full pressure range of investigation. Due to uncertainties involved in the determination of absolute values of $M$, we normalized all curves to $M(6.5 K) = 0$ and $M(1.8 K) = -1$.

As shown in Fig. \ref{fig5_MT} (a), when increasing pressure, the superconducting transition remains single and sharp up to 0.29 GPa. A sudden decrease of the onset transition temperature is observed between 0.29 GPa and 0.46 GPa. In the pressure region of 0.46 GPa to 1.01 GPa, the superconducting transition exhibits a double-transition feature. At 1.17 GPa, transition becomes single and sharp again. For decreasing pressure from 0.95 GPa to 0 GPa (see Fig. \ref{fig5_MT} (b)), all of the $M(T)$ curves exhibit double-transition features.

We summarize the $T_\text c$ values inferred from our magnetization measurements for BaBi$_3$ in Fig. \ref{fig6_phasediagram_MT} (a) in a $T-p$ phase diagram. To be consistent with $T^\text{offset}_\text c$ determined from resistance measurements, onset criteria of diamagnetism were used. In the case a double transition was observed, the individual onset temperatures were considered (see lower inset of Fig. \ref{fig5_MT} (a)).The directions of the triangle symbol of the data point indicate increasing ($\blacktriangleright$) and decreasing ($\blacktriangleleft$) pressure. The number associated with each data point represents the run number in the magnetization measurement. Letters $a$ and $b$ are used to label the two transition temperatures in case a double-transition feature was observed. As shown in Fig. \ref{fig6_phasediagram_MT} (a), three branches of $T_\text c(p)$ can be seen. Due to the abrupt change of $T_\text c(p)$ at $p\sim$ 0.3 GPa and $p\sim$ 0.9 GPa, we suggest three different superconducting phases existing under pressure which we will label in the following by ${\alpha}$, $\beta$ and $\gamma$. For each phase, over its range of stability, $T_\text c$ values are linearly suppressed by increasing pressure as shown in Fig. \ref{fig6_phasediagram_MT} (a). On increasing pressure, at low temperature, BaBi$_3$ starts with phase ${\alpha}$ at ambient pressure. When pressure is increased from 0.29 GPa to 0.46 GPa (Run 3 to Run 4), it enters an intermediate-pressure region (0.46 GPa $\leqslant p \leqslant$ 1.01 GPa, Run 4 to Run 6) where both features of phases $\beta$ and $\gamma$ are observed at low temperature. As pressure is further increased from 1.01 GPa to 1.17 GPa (Run 6 to Run 7), only phase $\gamma$ is observed. When decreasing pressure, BaBi$_3$ starts with pure phase $\gamma$ at 1.17 GPa. Decreasing pressure drives BaBi$_3$ again into a region (0.95 GPa $\geqslant p \geqslant$ 0.24 GPa, Run 8 to Run 12) where phases $\beta$ and $\gamma$ are observed. However, as indicated in the figure, further decreasing pressure does not restore the pure ${\alpha}$ phase as BaBi$_3$ starts with. Instead, a coexistence of phases ${\alpha}$ and $\beta$ is observed from 0.13 GPa to 0 GPa (Run 13 to Run 15). It should be noted that the phase diagram in Fig. \ref{fig6_phasediagram_MT} (a) is quite different from that shown in Ref. \onlinecite{Jha2017}, where $T_\text c$ is first increased upon increasing pressure up to 0.5 GPa with the rate of 1.22 K/GPa and then almost saturates at 0.75 GPa. This could be due to a combination of the relatively small data density, large pressure steps, possible hysterisis effects and not recognizing double-transition as mixture of phases.

To better demonstrate the phase transformation in the pressure regions where multiple phases are observed, we present in Fig. \ref{fig6_phasediagram_MT} (b) the relative phase portions of superconducting ${\alpha}$, $\beta$ and $\gamma$ as a function of pressure at low temperature. The relative phase portions for different phases are obtained by calculating the corresponding drop values $\Delta M$ (Indicated in the inset of Fig. \ref{fig5_MT} (a)) in the $M(T)$ data. The sizes of symbols in Figs. \ref{fig6_phasediagram_MT} (a) and (b) are roughly proportional to the relative phase portions. As shown in the figure, BaBi$_3$ starts with 100$\%$ ${\alpha}$ phase at ambient pressure. As pressure increases from 0.29 GPa to 0.46 GPa, the relative phase portion of ${\alpha}$ is entirely suppressed and phases $\beta$ and $\gamma$ emerge. $\beta$ is the majority phase in the mixture with almost 100$\%$ phase portion up to 0.51 GPa. Further increasing pressure suppresses the relative phase portion of $\beta$ and stabilizes $\gamma$ until phase portion of $\gamma$ reaches 100$\%$ at 1.17 GPa. For decreasing pressure, similar behavior of phase portions for $\beta$ and $\gamma$ is observed for the pressure region of 0.95 GPa to 0.24 GPa. From 0.13 GPa to 0 GPa, phase portion of $\beta$ decreases as phase portion of ${\alpha}$ increases, and BaBi$_3$ ends up with $\sim$ 75$\%$ of $\beta$ and $\sim$ 25$\%$ of ${\alpha}$ at 0 GPa. By both decreasing and increasing pressure in the magnetization measurement, figures \ref{fig6_phasediagram_MT} (a) and (b) clearly reveal that the transition between $\alpha$ to $\beta$ and the transition from $\beta$ to $\gamma$ are each first order. Figure. \ref{fig6_phasediagram_MT} (b) clearly shows wide pressure ranges of coexistence of $\alpha$ and $\beta$ as well as $\beta$ and $\gamma$. In addition, upon releasing pressure we find that $\beta$ phase can exist in a metastable phase.

To analyze the interrelation between superconductivity and the various high-temperature anomalies observed, $R(T,p)$ data were further analyzed. The corresponding surface-gradient magnitude $\sqrt{(\partial R/\partial T)^2+(\partial R/\partial p)^2}$ was calculated as a function of both temperature and pressure and is shown as color plot in Fig. \ref{fig7_phasediagram}, together with $T_\textrm S$, $T_\textrm S'$, $T_p$ and $T_\textrm c$ data that were obtained from both resistance and magnetization measurements. In Fig. \ref{fig7_phasediagram} (a), it is shown that $T_p$ is revealed as a sharp anomaly in the color plot. $T_\text S$ is revealed as a sharp anomaly in the color plot up to 1.05 GPa. The transition at $p\sim$ 0.9 GPa, $T_\text S'$, is revealed in the color plot as well, though more broadly. We suggest that $T_\text S$ and $T_\text S'$ lines are likely to be one transition line inferred from different criteria, as the color plot shows that they connect smoothly with each other. Figure \ref{fig7_phasediagram} (b) presents the blow up of the low-temperature region ($T=$ 4 K to 7 K). $T^\textrm{offset}_\textrm{c}$ from resistance measurement and $T^\textrm{onset}_\textrm{c}$ from magnetization measurement are plotted together for consistency. As shown in the figure, $T^\textrm{offset}_\textrm{c}$ is suppressed from 6 K to 5.8 K in the low pressure region (0 GPa $\leqslant p \leqslant$ 0.27 GPa), then it undergoes a sudden drop from 5.8 K to 5.3 K when entering the intermediate pressure region (0.33 GPa $\leqslant p \leqslant$ 1.05 GPa). In the intermediate pressure region, $T^\textrm{offset}_\textrm{c}$ is initially suppressed to 5 K by 0.71 GPa and then increases to 5.9 K at 0.93 GPa. At even higher pressures (1.16 GPa $\leqslant p \leqslant$ 2.13 GPa), $T^\textrm{offset}_\textrm{c}$ slowly decreases to 5.5 K. A subtle kink-like anomaly at 1.54 GPa is observed and will be discussed in detail later in the text below. $T^\textrm{onset}_\textrm{c}$ from magnetization measurement for increasing pressure matches very well with the $T^\textrm{offset}_\textrm{c}$ from resistance measurement. The multi-step transition pressure region for resistance measurement agrees with the double-transition pressure region for magnetization measurements. Importantly, the pressure region in which the double-transition is observed is also enclosed by the $T_p$, $T_\textrm S$ and $T_\text S'$ lines. Furthermore, the pressures where sudden changes in $T_\text c(p)$ are observed ($p\sim$ 0.3 GPa and $p\sim$ 0.9 GPa) coincide with the $T_p$ and $T_\text S'$ anomalies in the color plot. These observations demonstrate the strong interrelation between the superconductivity in BaBi$_3$ with the high-temperature phase transitions.

\begin{figure}
	\includegraphics[width=8.6cm]{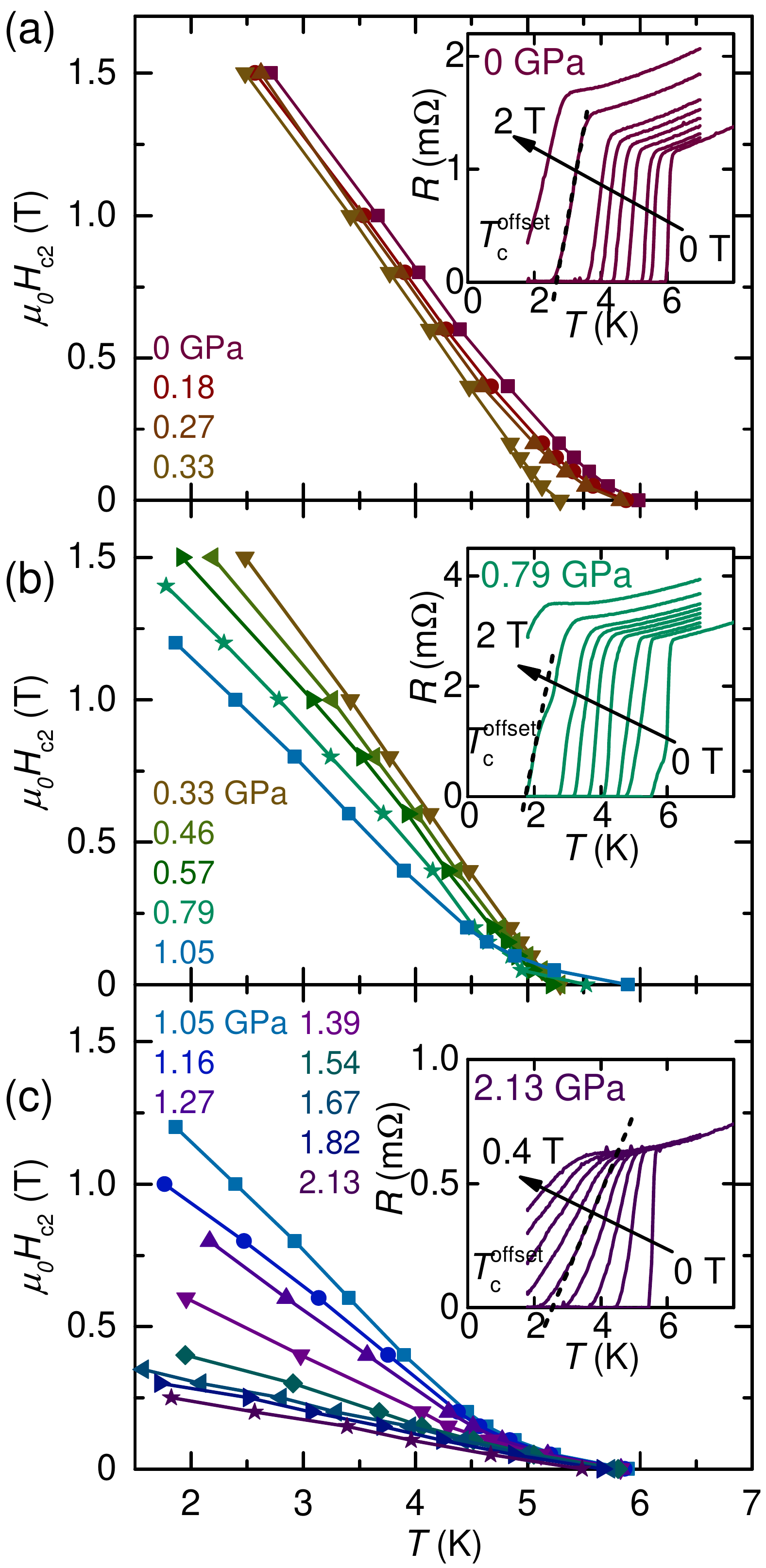}%
	\caption{Temperature dependence of the superconducting upper critical field $H_{c2}(T)$ for (a) $p\leq$ 0.33 GPa, (b) 0.33 GPa $\leq p \leq$ 1.05 GPa and (c) 1.05 GPa $\leq p \leq$ 2.13 GPa. $T^\textrm{offset}_\textrm{c}$ as shown in the insets is taken from resistance measurement. Insets show representative resistance data under magnetic fields.
		\label{fig8_hc2}}
\end{figure}

\begin{figure}
	\includegraphics[width=8.6cm]{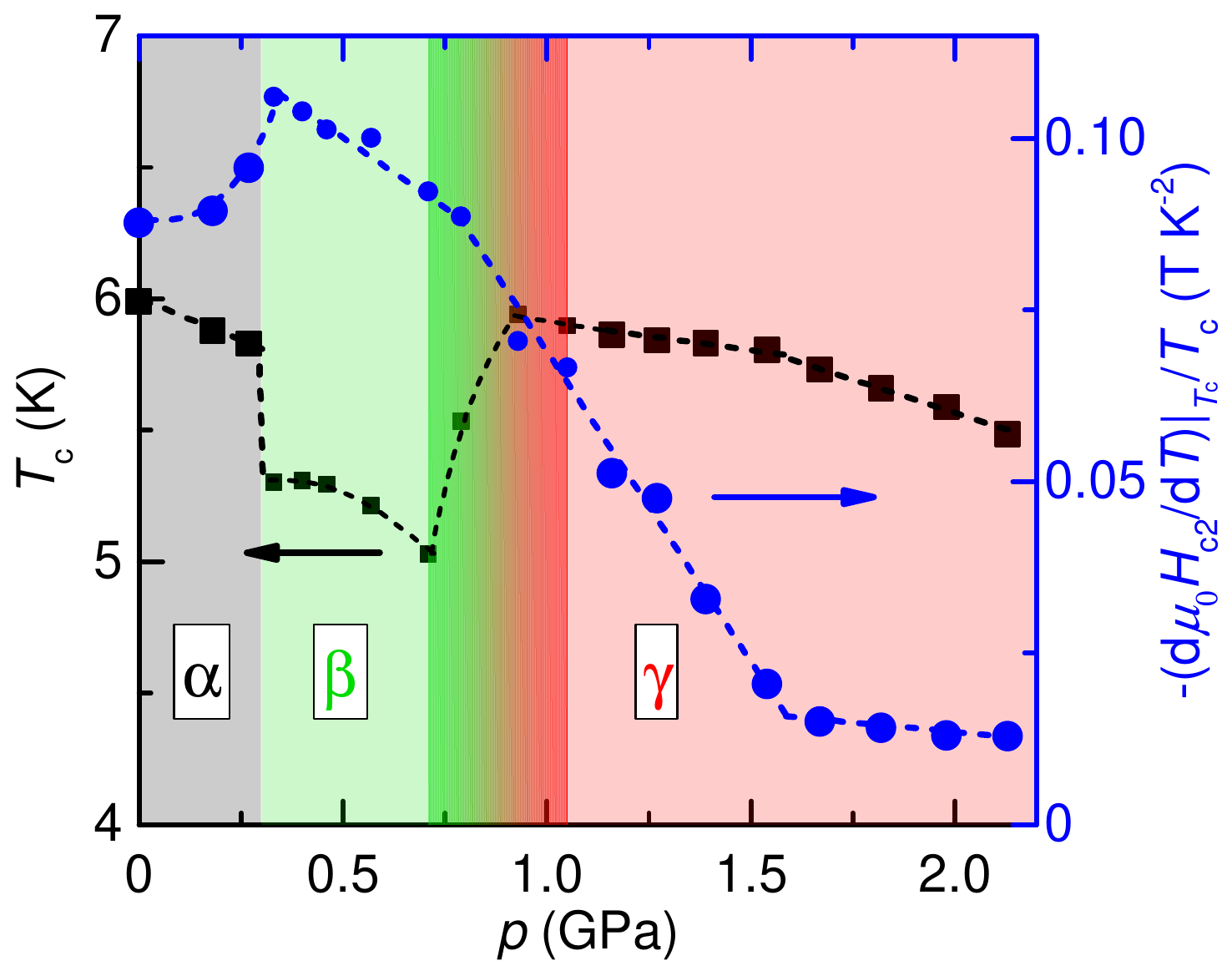}%
	\caption{Pressure dependence of the normalized upper critical field slope, -(1/$T_\textrm c$)($d\mu_oH_{\textrm c2}$/$dT$)$|_{T_\textrm c}$, plotted together with $T^\textrm{offset}_\textrm{c}$ from resistance measurement. Smaller symbols indicate the pressure range where superconducting transition shows multiple steps. Different phases ${\alpha}$, $\beta$ and $\gamma$ are proposed for different pressure regions as indicated in Fig. \ref{fig7_phasediagram}. Dashed lines are guidance to eyes.
		\label{fig9_norslope}}
\end{figure}

To further study the nature of the superconducting state in the $\alpha$, $\beta$ and $\gamma$ phases, we examined the response of superconductivity to external field. Figure \ref{fig8_hc2} shows the temperature dependence of the superconducting upper critical field $H_\text{c2}$ at various pressures. The insets show representative resistance data sets in the three pressure regions which were used to extract the data present in the main panels. As shown in the figure, for low- and high-pressure regions ($p\leqslant$ 0.33 GPa and $p\geqslant$ 1.16 GPa), the superconducting transition stays one single transition under magnetic fields. In contrast, for the intermediate-pressure region, the multiple-step nature of the superconducting transition persists in magnetic fields. For all of the pressures, $H_\text{c2}$ is linear in temperature except for low magnetic fields. The curvature at low fields has been observed in other superconductors and can be explained by multi-band nature of superconductivity \cite{Kogan2012,Kaluarachchi2016,Xiang2017PRB,Xiang2018PRB}, which is also the case BaBi$_3$ \cite{Haldolaarachchige2014,Shao2016}. The slope of the temperature-dependent $H_\text{c2}$ was obtained by linear fitting the $\mu_0 H_\text{c2} (T)$ data above the curvature (data above 0.2 T for low and intermediate pressure regions, data above 0.1 T for high pressure regions). Similar analysis was performed in literature for other superconductors, see Refs. \onlinecite{Taufour2014,Kaluarachchi2016,Xiang2017PRB, Xiang2018PRB}. Generally speaking, the slope of the upper critical field normalized by $T_\textrm{c}$, is related to the Fermi velocity and superconducting gap of the system\cite{Kogan2012}. In the clean limit, for a single-band,
\begin{equation}
-(1/T_\textrm c)(d\mu_oH_{\textrm c2}/dT)|_{T_\textrm c} \propto 1/v_F^2,
\label{eq:Hc2}
\end{equation}
where $v_F$ is the Fermi velocity. Even though the superconductivity in BaBi$_3$ is multiband\cite{Haldolaarachchige2014,Shao2016}, Eq. \ref{eq:Hc2} can give qualitative insight into changes induced by pressure.

Figure \ref{fig9_norslope} presents the pressure dependence of the normalized slope of the upper critical field, \hbox{-(1/$T_\textrm c$)($d\mu_oH_{\textrm c2}$/$dT$)$|_{T_\textrm c}$}, together with the $T^\textrm{offset}_\textrm{c}$ data. Smaller symbols indicate the intermediate-pressure region where the superconducting transition occurs in multiple steps. As shown in the figure, the normalized slope -(1/$T_\textrm c$)($d\mu_oH_{\textrm c2}$/$dT$)$|_{T_\textrm c}$, exhibits anomalies between 0.27 GPa and 0.33 GPa and between 0.75 GPa and 1.05 GPa, which coincides with the phase transition ranges for ${\alpha}$ to $\beta$ and $\beta$ to $\gamma$. Another anomaly is observed at 1.54 GPa, which coincides with the pressure where a small, kink-like anomaly in $T^\textrm{offset}_\textrm{c}$ occurs. Due to the absence of any feature in $R(T)$ at $T>T_\text c$ in this pressure region, we suggest that this small feature might be related to a change of band structure or to a Lifshitz transition, or some other change in $v_F$ within the $\gamma$ phase\cite{Lifshitz,Kang2015,Jo2016PRB,Xiang2017PRB,Xiang2018PRB}.

\begin{figure}
	\includegraphics[width=8.6cm]{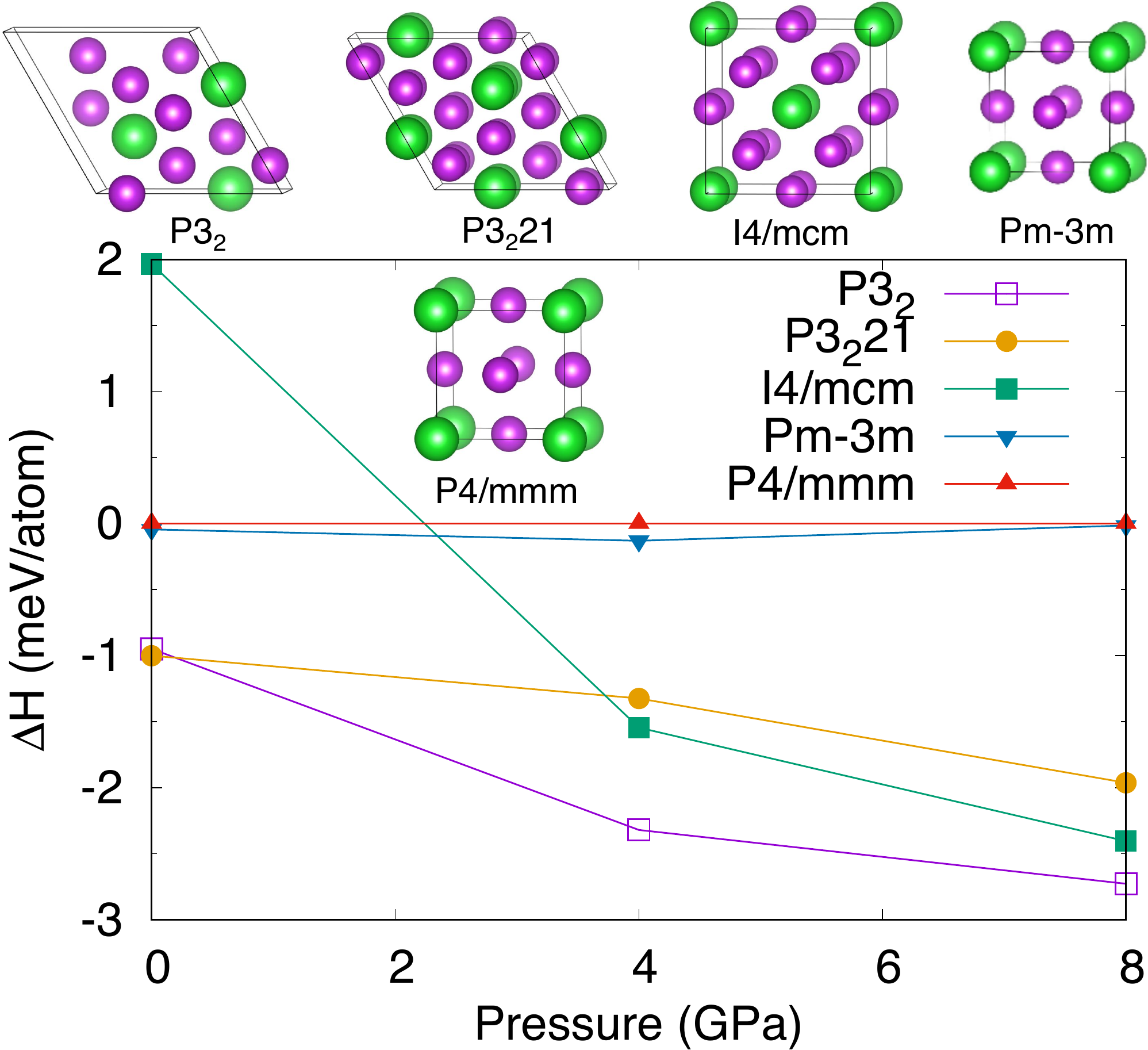}%
	\caption{Pressure dependence of relative formation enthalpies and crystal structures of low energy structures (see test for details). Green and purple balls are Ba and Bi atoms.
		\label{enthalpy_1}}
\end{figure}

Our studies show that BaBi$_3$ exhibits three different phases in a relatively small temperature and pressure range. The sudden changes in the superconducting character and the anomalies at high temperature suggest that structural degrees of freedom are crucial for understanding the behavior of BaBi$_3$ under pressure. Similar sudden change in $T_\text c(p)$ and associated high-temperature anomalies have been observed in PbTaSe$_2$ where a first-order structural phase transition is identified\cite{Kaluarachchi2017PRBa}. To gain insight to the pressure stabilized structures in BaBi$_3$, we performed first-principle calculations under pressure. Figure. \ref{enthalpy_1} presents the relative formation enthalpy (with respect to experimentally observed ambient-pressure $P4/mmm$ structure) at different pressures of the five lowest-energy structures found from our crystal structure search at zero temperature. The difference between $Pm-3m$ and $P4/mmm$ structures is the slight difference in lattice parameters $a$ and $c$ in $P4/mmm$ structure. The $I4/mcm$ structure could be viewed as distorted $P4/mmm$ structure where the Bi atoms are moved out of the face-centered position as can be seen in Fig. \ref{enthalpy_1}. The $P3_2$ and $P3_221$ structures can also be viewed as distorted $P4/mmm$ structure as well (not obvious in Fig. \ref{enthalpy_1} but from different directions of view) but with distortions in the both Ba and Bi positions.

We found several structures with very small relative formation enthalpies. We note that these small energy differences within 2 meV/atom could be within the error of DFT calculations and the pressure range in the DFT calculations is not exactly the same pressure range as in the experiment, but the trend of formation enthalpy change with pressure may be observed to speculate possible pressure stabilized structures in BaBi$_3$. As shown in the figure, the relative formation enthalpies of different structures response differently to external pressure, e.g., for $I4/mcm$ relative formation enthalpy decreases very fast at pressure between 0 and 4 GPa while that of $P3_221$ decreases slightly. At pressure larger than 4 GPa, 3 new structures are very competitive in formation enthalpy with the differences within 1 meV/atom and all of them more stable than $P4/mmm$ structure. Our crystal structure search and DFT calculation show that there are several structures very competitive in formation enthalpy and these structures are very likely more stable than ambient-pressure $P4/mmm$ structure under pressure. Experimental crystallographic data taken under pressure are needed to identify precisely which structures are the $\beta$ and/or $\gamma$ phases.

The transformations from $P4/mmm$ structure to $Pm-3m$ or $I4/mcm$ structures are minimal for the fact that only small changes in the structure need to be made as mentioned above. The response of $I4/mcm$ structure to pressure is the most robust among considered structures as mentioned above. On the other hand, the temperature and kinetics of transformation, which are not included in DFT calculations, also play important role in structural transformation. We may speculate that the $Pm-3m$ and $I4/mcm$ structure are the $\beta$ and $\gamma$ phases observed in experiment, respectively. The possibilities of $P3_2$ and/or $P3_221$ structures were observed in experiment are not exclusively eliminated though.

The calculation results are reasonable if we look into the ambient-pressure structural information for the $A$Bi$_3$ family. As mentioned in the introduction, at ambient-pressure and room temperature, BaBi$_3$ crystallizes in tetragonal structure with only a small difference in the $a$ and $c$ lattice parameters ($a$ = 5.06(1) {{\AA}} and $c$ = 5.13(2) {{\AA}}), which we label as ${\alpha}$ phase. In contrast, both of its neighboring compounds SrBi$_3$ and LaBi$_3$ crystallize in cubic structure with lattice parameters $a$ = 5.05(3) {{\AA}} and 4.99(2) {{\AA}}, respectively, i.e., with smaller lattice parameters and unit cell volumes. As applying hydrostatic pressure to BaBi$_3$ will decrease its lattice parameters, we assume that pressure tends to drive BaBi$_3$ to the higher-symmetry cubic structure, as realized in the neighboring compounds\cite{Kinjo2016,Jha2017,Leger1993,Venkateswaran1998,Bouvier2002,Einaga2011}. It is worth noting that the drop of $T_\text c$ at $T_p$ is consistent with the fact that SrBi$_3$ has lower $T_\text c$ than BaBi$_3$. With regard to the high pressure sensitivity of this compound, we would like to mention the results of previous calculations of the SOC influence on the phonon spectra of $A$Bi$_3$\cite{Shao2016}. Whereas for SrBi$_3$ these calculations indicate that the cubic structure is stable even without considering SOC, the same calculations find that SOC is necessary to stabilize the ambient-pressure tetragonal structure in BaBi$_3$. This demonstrates that the interplay of electronic and structural degrees of freedom in BaBi$_3$ places this material close to a structural instability. Together with the multiple phase transitions observed in the present work in a small pressure range, we establish BaBi$_3$ as a good platform to study the interplay of structure and superconductivity in the presence of spin-orbit coupling.

\section{Conclusion}

We establish three different phases ${\alpha}$, $\beta$ and $\gamma$ in BaBi$_3$ under pressure up to 2.13 GPa, each of which are superconducting at low temperatures. In the low-pressure region, BaBi$_3$ is purely in ${\alpha}$ phase for the whole investigated temperature range. When pressure is firstly increased at high temperature, BaBi$_3$ transfers into $\gamma$ phase through a likely first order transition. In $\gamma$ phase, by lowering temperature, the compound goes through another first-order transition to $\beta$ phase. Further increasing pressure suppresses the transition temperature of $\gamma$ to $\beta$ phase and in the high-pressure region, BaBi$_3$ stays in $\gamma$ phase for the whole investigated temperature range. Based on crystal structure search and DFT calculations, we speculate the phase transitions between ${\alpha}$, $\beta$ and $\gamma$ to be related to structural degrees of freedom. With ${\alpha}$ phase being the ambient-pressure tetragonal structure ($P4/mmm$), $\beta$ phase could be cubic ($Pm-3m$) and $\gamma$ phase could be distorted tetragonal structure ($I4/mcm$). Measurement of the superconducting upper critical field analysis exhibits an anomaly at $p=$ 1.54 GPa, suggesting a pressure-induced band structure change or Lifshitz transition within the $\gamma$ phase. Our results establish BaBi$_3$ as a good platform to study the interplay of structure and superconductivity in the presence of spin-orbit coupling.

\begin{acknowledgements}
	We would like to thank A. Kreyssig for robust (but not oaky) and enlightening discussions. This work is supported by the US DOE, Basic Energy Sciences, Materials Science and Engineering Division under contract No. DE-AC02-07CH11358. L. X. was supported, in part, by the W. M. Keck Foundation. R.A.R. was supported by the Gordon and Betty Moore Foundation EPiQS Initiative	(Grant No. GBMF4411) and by FAPESP (Grant No. 2011/19924-2).
\end{acknowledgements}

\clearpage

\bibliographystyle{apsrev4-1}
%
\end{document}